\documentclass[aps,twocolumn,showpacs,superscriptaddress]{revtex4}
\usepackage{amsmath}
\usepackage{amsfonts}
\usepackage{amssymb}

\usepackage{graphics,graphicx}
\usepackage{dcolumn,bm}

\newcommand{\cu}
{\affiliation{Department of Physics, University of Calcutta, 
92 Acharya Prafulla Chandra Road, Kolkata 700009, India.}}

\begin{document}

\title
{Quantum random walk : effect of quenching}

\author{Sanchari Goswami}%
\email[Email: ]{sg.phys.caluniv@gmail.com}
\cu
\author{Parongama Sen}%
\email[Email: ]{psphy@caluniv.ac.in}
\cu

\begin{abstract}

We study the effect of quenching on a discrete quantum random walk by removing a detector placed at 
a position $x_D$ abruptly at time $t_R$ from its path.
The results show that this may lead to an enhancement of the occurrence probability at $x_D$ 
provided the time of removal $t_R < t_{R}^{lim}$ where $t_{R}^{lim}$ scales as $x_D{^2}$. 
The ratio of the occurrence probabilities for a quenched walker ($t_R \neq 0$)  and free walker ($t_R =0$) 
shows that it scales as $1/t_R$ at large values 
of $t_R$ independent of $x_D$. On the other hand if $t_R$ is fixed 
this ratio varies as $x_{D}^{2}$ for small 
$x_D$. The results are compared to the classical case. We also calculate the correlations as functions of both time and position.
\end{abstract}

\maketitle

\section{Introduction}

Quenching phenomena has been a much studied topic in both classical and 
quantum systems in the recent past.  A quenching process can be broadly defined as one in which 
a certain quantity or condition of the system (e.g., temperature, magnetic field etc.) is changed 
in time from an initial value to a final value  in a certain manner. Quenching can be studied 
in several possible ways. In certain cases, slow quenching is 
applied to obtain the classical ground states of the system as in the case
of a spin glass system where there are many minima in the energy landscape, separated by 
barriers which maybe  overcome by quantum tunnelling \cite{arnab, book1}. On the 
other hand, in systems with a quantum critical point, nonequilibrium dynamics is studied by fast or slow quenching of 
relevant variables (usually a field or 
interaction which is made time dependent).
Examples include transverse Ising  and XY models \cite{book1, dutta, debanjan, sirshendu}. 
In ultracold atoms in an optical lattice 
fast quenching is applied by  shifting the 
position of the trap potential and studying its response
\cite{cold1,cold2}. In case of the transverse Ising or XY model, there is a 
deviation from the equilibrium state under quenching as the quantum 
critical point is crossed and the quantity of interest is the \textquoteleft defect\textquoteright, 
which is the amount of departure from the actual equilibrium value.

We consider here a discrete quantum random walk (QRW) 
\cite{aharonov,nayak,kempe,kiss}, completely different from a classical 
random walk \cite{chandra,book2,redner},
on which the effect of quenching is studied. The QRW is made 
unitary by coupling the translation with chirality or rotation.  The state
of the walker is expressed in the $|x\rangle |d\rangle$ basis, where $|x\rangle$
is the position (in real space) eigenstate and $|d\rangle$ is the
chirality eigen state (either “left” or “right). 

\section{Quenched Quantum Walk : Scheme and Measurements}

Usually, a quantum random walk is studied with two kinds of boundary conditions.
In the infinite walk (IW), there is no boundary and in the  semi-infinite 
walk (SIW) there is one absorbing boundary. Measurement-wise this signifies that 
there is no detector in the first case up to the time of observation, while 
there is a detector all the time in a particular position in the second. 

We consider the case when a detector initially placed at $x_D$ is removed 
suddenly at time $t_R$. This is termed a quenching phenomena as the presence of the detector 
and its subsequent removal may be compared to the presence of a time dependent transverse 
field in Ising model, XY model etc. 
We call this walk the quenched quantum walk (QQW).
 
Classically, even when there is an  absorbing boundary, a random walker will simply be either completely 
absorbed there, or if not, will propagate
like a free walker. In the quantum case on the other hand, the walk 
exists with non-zero probability at different locations and the absorption will 
thus take place with a probability. Experimentally, this means 
that if the particle is detected, the evolution will be stopped. 
But if not, the walker will survive with non-zero and modified probabilities at the 
other locations \cite{feynman} 
which is completely different from one which propagates without a detector \cite{goswami}.  



In the QRW, the position of the particle, $\psi({x},t)$ is given as,
\begin{equation}
\psi({x},t)=
  \begin{bmatrix}
    \psi_{L}({ x},t)\\
    \psi_{R}({ x},t) 
  \end{bmatrix}.
\end{equation}
\noindent
Here we have chosen the 
Hadamard coin \cite{nayak,amban} unitary operator $H$ to perform the rotation which is coupled to the 
translation. $H$ is given by 
\begin{equation}
H =\frac{1}{\sqrt{2}}
 \begin{bmatrix}
  1 & 1 \\
  1 & -1 
  \end{bmatrix}.
\end{equation}
The occupation probability of site $x$ at time $t$ is given by
$|\psi_{L}({ x},t)|^2 + |\psi_{R}({ x},t)|^2$
\noindent
; sum of these probabilities over all $x$ is $1$. 
 The walk is initialized at the origin with
$\psi_{L}(0,0) = a_0,  \psi_{R}(0,0)= b_0; ~~a_0^2 + b_0^2 =1.$
We have taken a symmetric 
walk; 
 $a_0=1/\sqrt{2}, b_0=i/\sqrt{2}$.


Let the detector be placed at some given
site $x_D$. In general, if the particle is at $x_D$ with probability $\alpha$ and the detector 
detects the particle with probability $\beta$, then the total absorption probability 
at $x_D$ will be $\alpha\beta$. In our case we have chosen $\beta = 1$. 
\begin{figure}
 \noindent \includegraphics[clip,width= 4.5cm, angle=0]{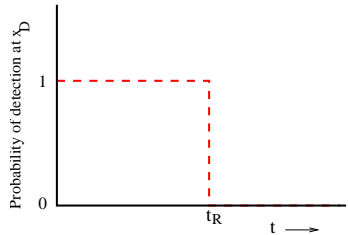}
\caption{(Color online)
The probability of the walker getting detected at a point $x_D$ is shown against time.
}
\label{pre}
\end{figure}
Removing the detector  at  time $t_R$ is equivalent to having a step function behaviour of the probability of being detected
at $x_D$ as a function of time as shown in Fig. \ref{pre}. 


At each time step when the ensemble is measured
the amplitudes describe only to the surviving copies. We may choose our scheme of measurement in 
two ways. Let the normalized occupation probability
at  $x$ at time $t$ be 
 denoted by $\tilde f(x,t)$. Thus $\tilde f(x, t)$  
 denotes the 
fraction of the copies that survived the measurement up to  time $t_R$  (not the
fraction of the initial population) 
which reaches $x$ at time $t$.
If the particle survives absorption, then $\tilde f(x,t)$ will be the conditional probability of finding 
it at $x$ at time $t$ in a \textquoteleft single\textquoteright observation. However, 
the average measure $f(x,t)$ takes into account the absorption probability and is given by 
$\tilde f(x,t) \times $ survival probability; $\sum_x \tilde f(x,t) = 1$ whereas 
$\sum_x f(x,t) = 1 -d$ where $d$ is the 
probability that it was absorbed earlier.
This issue 
of two types of measurement was already addressed in \cite{goswami}.

 

As $x_D$ and $t_R$ are the  parameters of the system, we further modify our notation: 
$\tilde  f(x, t, x_D, t_R)$ and 
 $f(x, t, x_D, t_R)$  are the normalised and average  occupation probability of site $x$ at 
time $t$ respectively given $x_{D}$ and $t_R$. 



Certain limiting cases can be immediately identified:

(i)  If $t_R=0$ then the normalised and average occupation probabilities become same and are identical to  
the usual occupation probability of a quantum random walker when there is no detector at all; the IW case.

(ii) If $t_R=\infty$, it is a case of SIW.

(iii) If $x_D>>0$, the walk will be IW-like in finite times.

The study for quench must be for a time $t \geq t_R \geq x_D$, otherwise 
the removal of the detector does not affect the probabilities. [Note that, here the numerical value of $x_{D}$ 
is considered while comparing with time $t$.]


\section{Results}

\begin{figure}
 \noindent \includegraphics[clip,width= 6cm, angle=0]{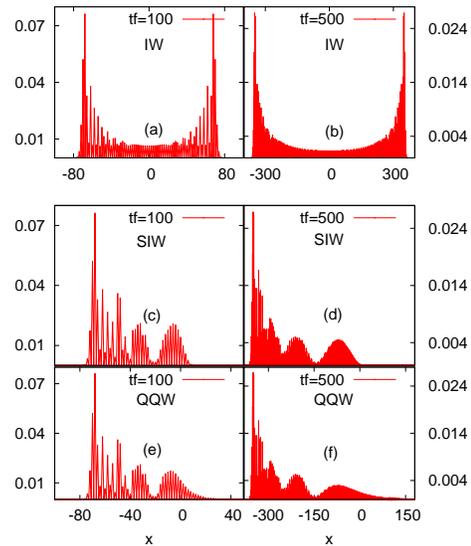}
\caption{(Color online)
Snapshots of the probabilities of occupation $f$ for a symmetric walker under 
different conditions are shown  for 
$t=100$ (left panel) and $500$ (right panel) against position $x$.  (a) and (b)  are results for   IW.  In (c) and (d) the results 
are shown for a SIW  when a  detector is placed at $x_{D} = 10$ for all times.  
In (e) and (f)  QQW results are shown where the detector is initially placed at $x_D = 10$ but is removed at time $t_R = 50$. 
}
\label{snapshots}
\end{figure}

We first present some snapshots of the probabilities of occupation $f(x)$ of different sites at 
different times for the IW, SIW and QQW 
 in Fig. \ref{snapshots} to get a comparative picture of the dynamics in the three cases. 
For a  quantum walker, the  displacement in time $t$ is proportional to $t$ and 
the maximum of the occupation probability occurs at a value of $x \sim t/\sqrt{2}$.
These features are clearly shown for the IW. For the SIW and QQW, the detector  placed at $x_D$ 
will make the probabilities different; up to $t = t_R$, QQW and SIW are equivalent.
However, for $t > t_R$, as shown in the figure, the probability ``spills out'' beyond $x_D$ for the QQW,
although far away from $x_D$, there is little difference. In fact away from the boundary, we find that
maximum probability is again at a value of $|x| \sim t/\sqrt{2}$.

\begin{figure}
\includegraphics[width= 4cm, angle=270]{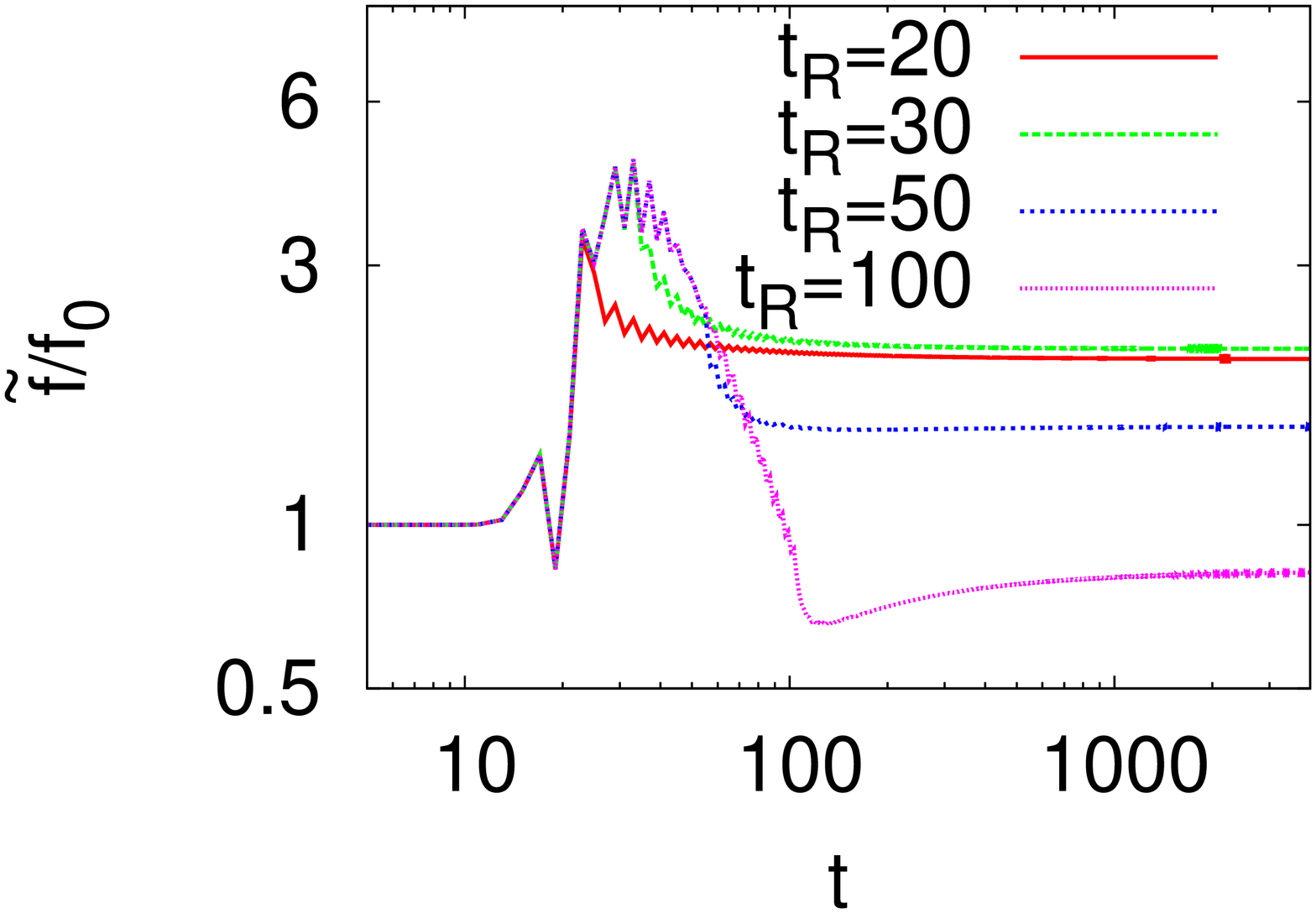}
\includegraphics[width= 4cm, angle=270]{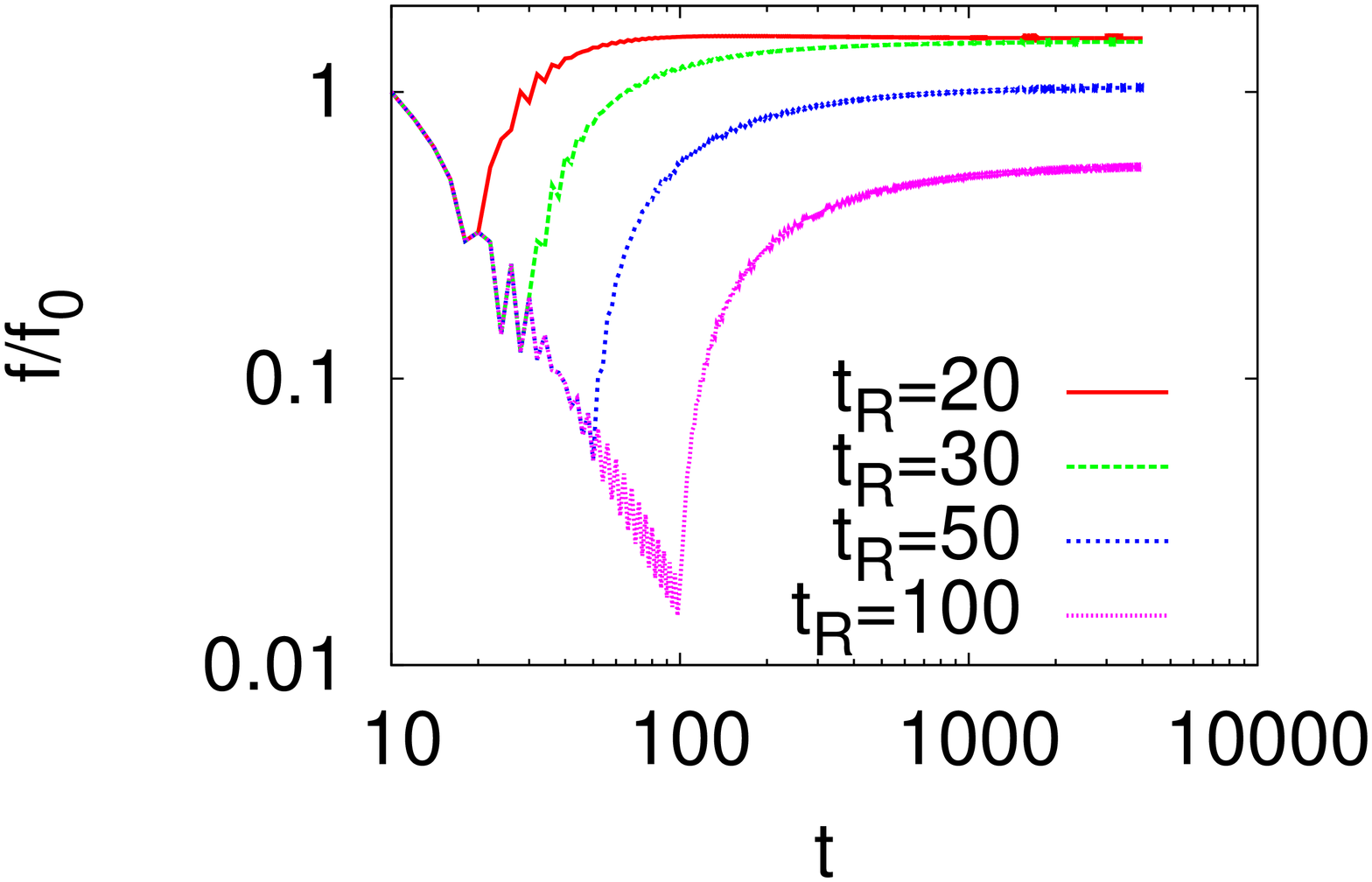}
\caption{(Color online) Top: Ratio of $\tilde f(x_{D}, t, x_{D}, t_{R})$ and $f_0$ for $x_{D}=10$ and $t_{R} = 20, 
30, 50, 100$. Bottom: Ratio of $f(x_{D}, t, x_{D}, t_{R})$ and $f_{0}$ for $x_{D}=10$ 
for the same values of $x_D$ and $t_R$.
}
\label{compare}
\end{figure}

We next present the data for a fixed value of $x=x_D$ at different times $t$
for both $\tilde f(x,t,x_{D},t_{R})$ and $f(x,t,x_{D},t_{R})$, using 
the shorthand notation $\tilde f$ and $f$ for these two quantities respectively.
To study the effect of quenching,   the ratios of the probabilities at different positions and times for the IW and the QQW 
may be calculated for given values of $x_D$ and $t_R$. This is in tune with the 
measure of defects in studies of quenching in the quantum spin models.
The ratios  
$\tilde f/f_0$  and $f/f_0$ are computed where $f_0$  is the occupation probability for the IW.
$\tilde f/f_0$ shows that it can attain values much larger than 1 for a short time $t > t_R$ 
before      
saturating at larger times to values larger or smaller than $1$ depending on $t_R$. These are shown in 
Fig. \ref{compare}.

\begin{figure}
\includegraphics[width= 4.5cm, angle=270]{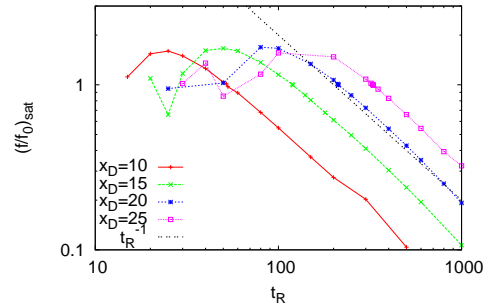}
\caption{(Color online) Plots of $(f/f_{0})_{sat}$  against $t_{R}$ for $x_{D}=10,15,20,25$ respectively. For 
large $t_{R}$ the log-log plot shows a slope of $-1$.
}
\label{sat_value}
\end{figure}

 The ratio of 
$f/f_{0}$ are also 
clearly different from $1$ beyond  $t=t_{R}$. However, it shows a much more
regular behaviour compared to $\tilde f/f_0$ although it takes a longer time to saturate (Fig. \ref{compare}). 
As the detector is placed at $x_D$, up to $t=x_D$, the walker is not affected 
by the presence of the detector and the ratios are equal to $1$ for all $t_R$.  
Beyond $x_D < t < t_{R}$,  the initial population of the ensemble decreases 
due to detection at site $x_D$. As time goes on, larger fraction of the initial population get lost and 
the ratio thus decreases gradually. 
Then as the detector is removed,  the probability grows at $x_D$, such that the ratio starts increasing
before reaching the saturation value. In fact the results are of interest mainly at $t > t_R$, after the 
detector is removed, when the QQW is different from the SIW and the ratio starts growing. Interestingly, contrary to naive 
expectation, the ratio may saturate 
at values larger than unity for small $t_R$. The explanation for $(f/f_{0})_{sat}$ 
having value $> 1$ is given later in the paper after we present the results for $x \neq x_{D}$.
The saturation values of the ratio, $ (f/f_0)_{sat}$ plotted against 
$t_R$ shows an initial non-monotonic
behaviour, but for large $t_R$, $(f/f_0)_{sat}$ clearly scales as $1/t_R$ for all values of $x_D$ (Fig. \ref{sat_value}).

\begin{figure}
\includegraphics[width= 4.5cm, angle=270]{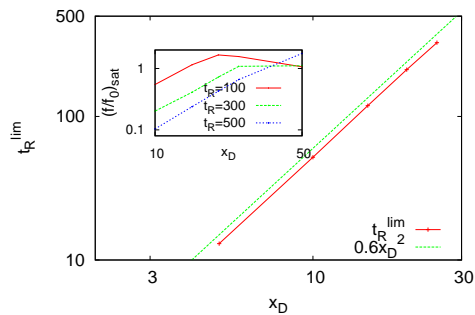}
\caption{(Color online) Plot of $t_{R}^{lim}$  against $x_D$. The log-log plot shows a slope of $2$. In the inset 
$(f/f_{0})_{sat}$  against $x_D$ is plotted for different $t_{R}$.
}
\label{cross1}
\end{figure}

From Fig. \ref{sat_value}, we note that there is a 
value of $t_R$ beyond which $(f/f_0)_{sat}$ is never $> 1$, i.e., beyond this particular 
$t_R = t_{R}^{lim}$, $(f/f_{0})_{sat}$ monotonically 
decreases below $1$. We find that $t_{R}^{lim}$ 
varies as $x_{D}^{2}$ (Fig. \ref{cross1}). Moreover, for fixed $t_R$, the  saturation value of the ratios
show a variation with $x_D$: for large $t_R$, $(f/f_{0})_{sat}$ varies as $x{_D}^2$, for smaller values of $t_R$,
this variation is valid over a small range of $x_D$. Combining the above two results, we conclude, for large $t_R$, 
\begin{equation}
 (f/f_{0})_{sat} = k x{_D}^2/t_R
\label{scaling_form}
\end{equation}
where $k$ is a constant with dimension of inverse length. That $(f/f_{0})_{sat}$ should decrease with $t_{R}$ is 
expected as the probability of detection increases with larger $t_{R}$. On the other hand as $x_{D}$ is increased, 
the walk remains unaffected for longer times which implies that $(f/f_{0})_{sat}$ 
should increase with $x_{D}$ ($x_{D} \rightarrow \infty$ makes 
$f=f_{0}$). However, the exact scaling form eq. (\ref{scaling_form}) is not obvious.

Before discussing other issues it is worthwhile to compare the results of the 
quantum case to the classical quenched random 
walk (CQW)  under identical fast quenching. For a classical random walker, the occupation 
probability $f_{c}(x,t)$ is simply given by $f_{c0}(x,t) \times $survival probability, 
where $f_{c0}(x,t)$ is the occupation probability for the classical walker in absence of 
any boundary. The survival probability is given by $1- \int_{0}^{t_{R}} F_{c}(x_{D},t) dt$, 
where $F_{c}(x_{D},t)$ is the first passage probability at $x_{D}$ at time $t$. 
So for the classical case, 
\begin{equation}
 \frac{f_{c}}{f_{c0}} = 1- \int_{0}^{t_{R}} F_{c}(x_{D},t) dt ,
\label{ratio}
\end{equation}
and is always less than $1$.
Moreover, it is independent of $x$ and $t$. $f_{c}/f_{c0}$ for $t_{R} \rightarrow \infty$ scales as 
$\frac{x_D}{\sqrt{t_{R}}}$ \cite{redner}. 
For the quantum case in contrast $(f/f_{0})_{sat}$ has a different scaling behaviour 
with $x_D/\sqrt{t_R}$ as already shown.

In fact for a CQW, the ratio is simply identical to the persistence probability at time $t_R$, i.e.,
the probability that the walker has not visited the site $x_D$ till time $t_R$.
 For the  quantum walker,   the corresponding 
persistence probability is proportional to $\frac{1}{t{_R}^2}$ \cite{goswami}. 
Hence the saturation ratio is not linearly dependent on the  persistence probability 
in case of the QQW  in contrast to the CQW.

\begin{figure}
\includegraphics[width= 4.5cm, angle=270]{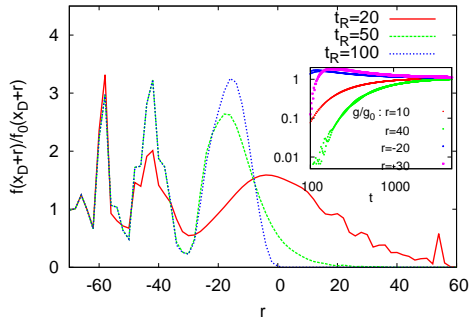}
\caption{(Color online) Ratio of the occupation probabilities $f(x_{D}+r)/f_{0}(x_{D}+r)$ 
against $r$ for $x_D = 10$ and $t_R=20, 50, 100$ at a particular time $t=100$. Inset shows $g/g_{0}$ 
against $t$ for $t_{R}=50$ for different $r$.
}
\label{comp}
\end{figure}

 Let 
us now discuss the results for  sites $x \neq x_{D}$; 
for a site separated by a distance $r$ from $x_D$ the occupation probability can be written as 
$f(x_{D}+r,t, x_{D},t_{R})$.
The ratio of the occupation probabilities $f(x_{D}+r)/f_{0}(x_{D}+r)$
 (in short hand notation)  
shows remarkably different behaviour for $r > 0$ and $r < 0$
at a finite value of $t$ (Fig. \ref{comp}).
For $r > 0$, the ratio 
goes to zero at a finite value of $r$, the  decay is smooth for large values of $t_R$ while for small $t_R$,
the decay is accompanied by small oscillations. 
For $r < 0$, the ratio shows an irregular behaviour with $r$, several peaks occur with the peak 
values much greater than $1$. However, looking carefully at Fig. \ref{snapshots},
it is evident that for sufficiently large $|r|$, when $r < 0$, the QQW and SIW behave in the same way,
and the ratio is not much affected by the removal of the boundary. 
Hence one can say that  the memory effects are strong here.

As function of time, the product of $\frac{f(x_{D}+r)}{f_{0}(x_{D}+r)}$ and 
$\frac{f(x_{D})}{f_{0}(x_{D})}$ are plotted in the inset of Fig. \ref{comp} for $x_D = 10$ and $t_R = 50$.  
The product approaches unity for all $r$, although for $r > 0$ it reaches unity from below 
and for $r < 0$ from above. this product can be identified as the ratio of two 
correlation functions $g(x_{D}+r,t,x_{D},t_{R})$ and $g_{0}(x_{D}+r,t,x_{D},t_{R})$ where
\begin{equation}
 g(x_{D}+r,t,x_{D},t_{R})=f(x_{D}+r,t,x_{D},t_{R})f(x_{D},t,x_{D},t_{R}) 
\label{corr_g}
\end{equation}
 and $ g_{0}(x_{D}+r,t,x_{D},t_{R})$ is obtained by replacing $f$ by $f_0$ in eq.(\ref{corr_g}).
Thus we note that the ratio of the correlation functions approach unity irrespective of the value of  $r$
for finite values of $t_R$.

The ratio of the two correlation functions may also be estimated for the classical case. Using the 
result of eq.(\ref{ratio}), $\frac{g_{c}}{g_{c0}}$  is simply the 
square of the persistence probability and is less than $1$ always irrespective of the value of $r$ and $t$. 
This is clearly different from the 
quantum case where the ratio goes to $1$ for all $r$ at large $t$. Moreover, in the quantum case there is a  
time dependence which is strongly dependent on the sign of $r$ at small $t$.



\section{Summary and Discussions}

In summary, we have studied quenching 
in a quantum system in a completely different sense compared to earlier works. 
The  observation that the occurrence probability of a QRW may actually be
enhanced by quenching is one of the main results of this study.
This is  a purely quantum mechanical effect.
Having the detector up to time $t_R$  means the quenched walker
cannot go beyond $x_D$, and the undetected
walker will move away from $x_D$. However, at a later time $t> t_R$, when the walker is free 
once again,
it can move towards  $x_D$ and go beyond. From Fig. \ref{snapshots}, it can also be seen 
that that most of the
contributions to $x_D$ and beyond come from the density of walkers closer to it,
as the occurrence probability far way from $x_D$ are not much affected with the
removal of the detector. At later times after the removal of the detector, the occupation probability profile 
approaches the IW picture as the local hill like structures smoothen out. This happens 
closest to $x_D$ at earlier times and slowly the further parts are affected. 
In comparison, in the infinite walk case, the walker has moved
reasonably  away from $x_D$ at $t_R$, such that the ratio can be greater than unity close to $x_D$. 
But when $t_R$ is greater than $t{_R}^{lim}$,
the ratio  can no longer exceed unity. 
At even later times,
the ratio saturates as the ``memory'' of the
detector gets erased in time.
On the other hand we find that memory effects are strong for $x << x_D$ where the removal of the detector is
more or less irrelevant.
Hence the effect of quenching is rather local.

Other important results  are the scaling behaviour of the quantities
like ${t_R}^{lim}$ with $x_D$ and $(f/f_0)_{sat}$ at $x_D$ as a function of $x_D$
or $t_R$. Although for QRW, the displacement varies linearly with time,
we find that the timescale ${t_R}^{lim}$ varies with $x_D$ in a quadratic manner.
The scaling behaviour of $(f/f_0)_{sat}$ is also drastically
different from the classical case.

The present work can be extended in many ways like making the probability of detection
dependent on time in a different way and slow quenching can be studied
when it decays algebraically. 
One can also, instead of having
a detector from time $t=0$ to $t=t_R$,  manipulate both the times at which
the detector is placed  and subsequently removed.
Quantum random walks of correlated particles have been shown to encode information \cite{photon} and
 quantum walk is capable of universal quantum computation \cite{childs}.
Quenching of quantum random walks may lead to some new features  in such contexts,
although it is too early to predict exactly how. 

\begin{acknowledgements}
The authors thank Arnab Das for some useful comments on the manuscript. 
SG acknowledges financial support from CSIR (Grant no. 09/028(0762)/2010-EMR-I). 
PS acknowledges financial support from DST (Grant no.SR-S2/CMP-56/2007)
\end{acknowledgements}

\end{document}